\documentclass[aps,prl,twocolumn,groupedaddress]{revtex4}
\usepackage{graphicx,latexsym}

\begin{document}

\title{Experimental evidence for a large critical transverse
depinning force in weakly-pinned vortices}

\author{J. Lefebvre, M. Hilke, and Z. Altounian}

\affiliation{ Dpt. of Physics, McGill University, Montr\'eal, Canada
H3A 2T8.}

\begin{abstract}
We present experimental evidence for the existence of a large
critical transverse depinning force. These results are obtained in
the weakly-pinned superconducting metal glasses
Fe$_{x}$Ni$_{1-x}$Zr$_{2}$ using crossed ac and dc driving currents.
We study the depinning force due to the transverse ac drive as a
function of a longitudinal dc drive. The ac/dc combination allows us
to separate the transverse drive from the longitudinal one. We show
that the force required for depinning in the transverse direction is
enhanced by a longitudinal drive, which leads to the existence of a
large transverse critical force.
\end{abstract}

\maketitle

The vortex state of type II superconductors is characterized by a
wealth of interaction phenomena: Whereas vortex-vortex repulsion
tends to order the system, thermal fluctuations and pinning from
material inhomogeneities introduce disorder in the vortex lattice.
This sort of competition between ordering and disordering makes the
vortex state rich in both static and dynamic phase transitions, as
well as in nonequilibrium phenomena.  While the effect of disorder
on the static case has been widely studied in the past years
\cite{NattermannPRL64, BouchaudPRB46, BlatterRMP66, GiamarchiPRL72,
GiamarchiPRB52} the driven case has still much to reveal.

A large number of studies have demonstrated that at high driving forces, a
disordered system will show ordering (dynamical
ordering) \cite{BhattacharyaPRL70, YaronPRL73, HellerqvistPRL76,
MarchevskyPRL78, ShiPRL67, MoonPRL77, RyuPRL77, OlsonPRL81, SpencerPRB55,
KoshelevPRL73, GiamarchiPRL76, BalentsPRL78, BalentsPRB57, LeDoussalPRB57,
PardoNature396, FangohrPRB63, FangohrPRB64, MarleyPRL74}. Experimentally, the crossover to
a more ordered vortex phase at large driving current is deduced in transport
measurements from the presence of a peak in the differential
resistance \cite{BhattacharyaPRL70, HellerqvistPRL76}, or from a decrease of
the low frequency broadband noise \cite{MarleyPRL74}; an increase of the
longitudinal correlation length in neutron diffraction experiments
has also revealed the existence of this dynamical ordering
phenomena \cite{YaronPRL73}. In addition, dynamic ordering was directly observed in magnetic
decoration experiments
\cite{MarchevskyPRL78, PardoNature396}.  Numerically and analytically, the
establishment of the existence of such dynamical phase transitions and
ordering has lead to the prediction of the existence of static channels in
which the vortices flow; these channels may be decoupled, in which case the
vortex phase obtained is called the moving transverse glass (MTG) and has
smectic order, or they may be coupled and one has the moving Bragg glass
(MBG).  It is predicted that these channels act as strong barriers against
transverse depinning, resulting in the existence of a finite transverse critical
force \cite{MoonPRL77, OlsonPRL81, GiamarchiPRL76, LeDoussalPRB57, OlsonPRB61, RyuPRL77, FangohrPRB63}.
Experimentally, the existence of this critical force has yet to be proved, and
the vortex channels have only been observed in magnetic decoration
experiments \cite{MarchevskyPRL78, PardoNature396} and STM
images \cite{TroyanovskiNat399}.

Here we study experimentally the transverse dynamics of vortices and
demonstrate the existence of a large transverse critical force. (Experimentally, the critical force is equivalent to the depinning force, so we will use the terminology ``critical force'' throughout the text to mean ``depinning force''.) We
find that for a system driven longitudinally with a dc current,
application of a small transverse force, provided by an ac current,
does not result in immediate transverse depinning. In some regimes,
the transverse force required for depinning the vortices in the
transverse direction is even increased by more than 30 \% with
respect to the force required in the longitudinal case, thus
implying the appearance of very strong barriers against transverse
motion. Numerical studies have found the ratio of the critical
transverse force over the critical longitudinal force
$\frac{f_{y}^{c}}{f_{x}^{c}}$ to be of the order of 1 \%
\cite{MoonPRL77, RyuPRL77, OlsonPRB61} or 10 \% \cite{FangohrPRB63}.
Following Ref.\cite{FangohrPRB63}, this ratio is expected to
increase for weaker pinning. However, finite size effects in
numerical simulations do not allow for studies in the limit of very
weak pinning, which is the regime of our experiments. Indeed, in our
experiments the vortex pinning is at least six times smaller and we
obtain a ratio $\frac{f_{y}^{c}}{f_{x}^{c}}$, which can exceed
100\%.

The measurements were performed on different samples of the metal
glass Fe$_{x}$Ni$_{1-x}$Zr$_{2}$ prepared by melt-spinning
\cite{AltounianPRB49} and which become superconducting below about
$2.4~$K depending on the iron content.  These samples are
particularly clean such that vortices are very weakly-pinned, and
pinning is isotropic and has no long-range order because of the
amorphous nature of the samples.  The samples have a very small
critical current density ($J_{c}\leq0.4~A/cm^{2}$), which means that
we can conveniently study the depinning mechanisms using a very
small driving current without introducing uncertainties due to the
heating of the sample related to the use of a large driving current.
The average vortex velocity can then be measured from the voltage
they produce perpendicular to their motion. These material were
found to be strong type II low temperature superconductors
\cite{HilkePRL91} from estimates of the different characterizing
length scales using standard expressions for superconductors in the
dirty limit \cite{KesPRB28}. These samples show a variety of phases
of longitudinal and transverse vortex motion, including a MBG-like
phase \cite{HilkePRL91, LefebvrePRB}, and hence are ideal for the
study of transverse depinning.

\begin{figure}
[ptb]
\begin{center}
\includegraphics[
natheight=7.654400in,
natwidth=6.285500in,
height=4.2211in,
width=3.4714in
]%
{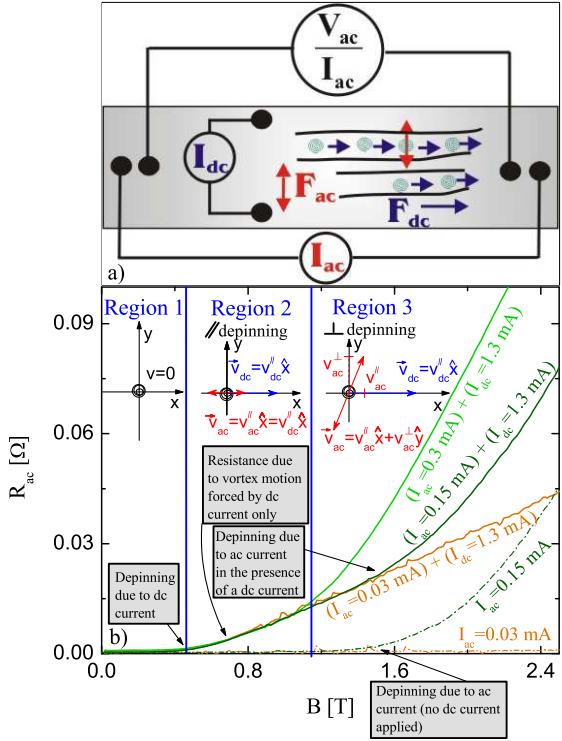}%
\caption{a) Drawing showing the contact configuration and resulting directions
of vortex motion. b) Resistance vs magnetic field measured with different
$I_{ac}$ and $I_{dc}$. \ The drawings show the trajectory followed by vortices
in the three regimes of vortex motion.}%
\label{RvsB}%
\end{center}
\end{figure}

We proceed by cooling the samples in a He$^{3}$ system to a temperature below
$0.4~$K.  We use a dc current as the longitudinal drive, and a 17 Hz ac
current provided by a resistance bridge as the transverse drive.  The
resistance is measured in the transverse direction with the resistance
bridge.  Indium contacts are soldered to the sample in the configuration
shown in Fig.(\ref{RvsB}a).  In a magnetic
field perpendicular to the sample plane, the force exerted on the vortices by
the dc current $I_{dc}$ applied along the short edge of the sample acts in the
direction $\vec{F}_{dc}=\vec{J}_{dc}\times\vec{\Phi}$, such that the vortices
move under the action of that force along the long edge of the sample.
 Similarly, because the ac current $I_{ac}$ is applied along the long edge of the
sample, the force it provides acts on the vortices in such a way
that they are oscillating along the direction of the short edge of
the sample. Therefore, in this configuration, the channels of
vortices are set up by the dc current in the longitudinal direction
along the long edge of the sample, and the transverse force is
provided by the ac current and directed along the short edge of the
sample.
 Evidently, the two sets of contacts used for dc driving and ac driving are
not perfectly perpendicular to each other, and the transverse
voltage measured also contains a component resulting from the ac
component along the dc longitudinally driven motion. This contact
misalignment $\alpha$ can be estimated to be of the order of
2${{}^\circ} $ for this particular sample and can be excluded
following the discussion below.  In all the figures, the error bars are found to be smaller than the size of the dot.

For instance, in Fig.(\ref{RvsB}b) we show measurements of the
transverse ac resistance as a function of magnetic field for zero
and non-zero longitudinal dc currents. This allows us to distinguish
three regions corresponding to three regimes of vortex motion:
Region 1 is characterized by vortices pinned in both directions, as
none of the currents is strong enough to depin the vortices, leading
to zero resistance. In Region 2, for a longitudinal dc current of
$I_{dc}=1.3~mA$, which is above the longitudinal depinning current of 0.55 mA,
we also measure an ac resistance, which is due to the small ac
component proportional to $\sin (\alpha)$ along the longitudinal
direction. In this region, where the depinning is only longitudinal,
the ac resistance is indeed independent of the transverse ac
current, clearly demonstrating that the vortices are pinned in the
transverse direction, since depinning is associated with strong
non-linearities. This is in stark contrast to region 3, where the ac
transverse resistance depends on the transverse ac current and
indicates the region where the vortices also start moving in the
transverse direction. The transverse depinning current is then
easily identified as the point in field and ac current where the ac
resistance depends on the transverse ac current. Hence, for a given
longitudinal dc drive, the pure transverse dynamics can be obtained
by subtracting the contribution due to a very small transverse ac
current with the same longitudinal dc drive, i.e., subtracting
$R_{ac}((I_{ac}=0.03mA)+(I_{dc}=1.3mA))$ from
$R_{ac}((I_{ac}=0.3mA)+(I_{dc}=1.3mA))$ in Fig.(\ref{RvsB}).

\begin{figure}
[ptb]
\begin{center}
\includegraphics[
natheight=5.260700in,
natwidth=5.904900in,
height=3.3364in,
width=3.7421in
]%
{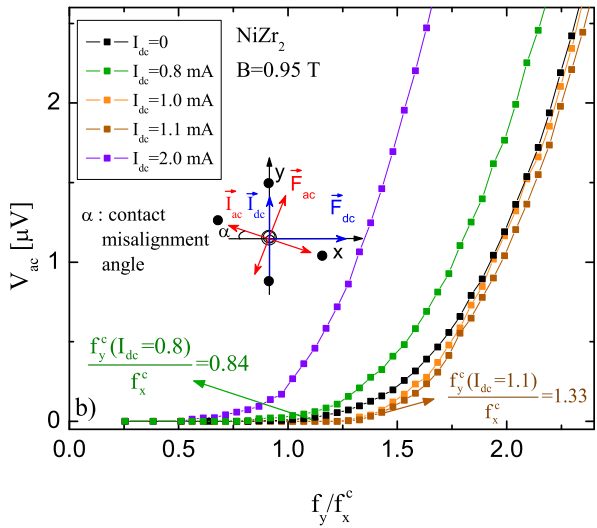}%
\caption{Transverse ac voltage versus applied ac force for different dc driven
cases normalized with the critical force in the static case. \ Inset: Drawing
showing the direction of the applied currents and resulting forces.}%
\label{VvsFratio}%
\end{center}
\end{figure}

\begin{figure}
[ptb]
\begin{center}
\includegraphics[
trim=0.327102in 0.000000in -0.516314in 0.000000in,
natheight=5.189700in,
natwidth=6.183400in,
height=3.0528in,
width=3.7421in
]%
{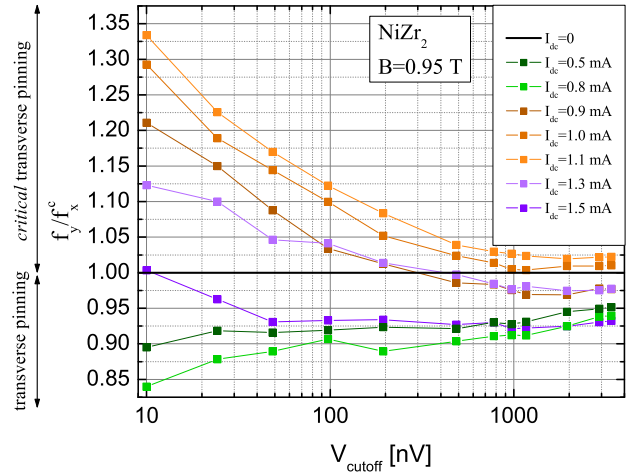}%
\caption{Ratio of the critical force in the driven case to that in the static
case as a function of transverse vortex velocity for different dc currents at
$B=0.95$ T.}%
\label{fratiovsvac}%
\end{center}
\end{figure}

Using the methodology described above, we show in
Fig.(\ref{VvsFratio}) the corrected transverse voltage versus the
transverse driving force for different longitudinal dc drives at a
magnetic field of $B=0.95~T$. The arrows in this figure indicate the
transverse depinning driving force normalized by the critical
longitudinal force ($\frac{f_{y}^{c}}{f_{x}^{c}}$) for two different
longitudinal drives using a voltage cutoff of 10 nV as depinning
threshold. The critical longitudinal force is obtained from the
transverse ac depinning current when the longitudinal current is set
to zero. This is justified since our system is isotropic in both
directions and this choice avoids errors due to the geometrical
factor when computing the difference in longitudinal and transverse
current densities. In addition, the longitudinal depinning currents
in the ac and dc driving case are very similar for these systems
\cite{LefebvrePRB}. From Fig.(\ref{VvsFratio}) we see that for
$I_{dc}=0.8~mA$ the transverse depinning force is slightly decreased
compared to the longitudinal one, in contrast to $1.0~mA\leq
I_{dc}\leq1.1~mA$, where the transverse depinning force is
increased, with a ratio $\frac{f_{y}^{c}}{f_{x}^{c}}$ reaching
$1.33$. Hence, in this range of longitudinal drives, strong barriers
against transverse motion are set up. The overall dependance can be
interpreted in the following manner: for
$I_{dc}=0.8~mA$, the force provided by this combination of magnetic field and current is just strong enough to cause depinning in the longitudinal direction, but the strength of the barriers
created by the channeling effect is still too small to cause a
strong pinning action in the transverse direction.  For $0.9~mA\leq
I_{dc}\leq1.1~mA$, the dc force does work very well at restraining
transverse vortex motion, and the force required to induce
transverse depinning even increases with the dc current (see Fig.(\ref{fratiovsvac})).  For
$I_{dc}=1.3~mA$, the ratio of the critical forces starts to decay;
the decay gets stronger for $I_{dc}\geq1.5~mA$.  This strong decay
is likely due to additional dynamic disorder, which could weaken the
barriers against transverse vortex motion \cite{FangohrPRB63}.

%EndExpansion

From Fig.(\ref{VvsFratio}) we extract the evolution of
$\frac{f_{y}}{f_{x}^{c}}$ for different transverse cutoff voltages.
The results are shown in Fig.(\ref{fratiovsvac}) for different
longitudinal driving currents, where the ratio of the critical
forces approaches one for transverse vortex velocities exceeding the
longitudinal vortex velocities. In fact, we obtain this ratio to be
exactly one when the transverse force equals approximately ten times
the longitudinal force (data not shown here). However, the fact that
this ratio does not reach one right after transverse depinning
occurs implies that the barriers against transverse vortex motion
not only delay transverse depinning, but also constrain transverse
vortex motion at larger velocities as well. This effect was also
observed numerically in Ref.\cite{FangohrPRB63}. More importantly,
this figure confirms the criticality of the transverse depinning
transition observed with the 0.9 mA $\leq I_{dc}\leq$1.5 mA
longitudinal drives. Indeed, in this regime, extrapolation of
$\frac{f_{y}}{f_{x}^{c}}$ to $V_{cutoff}=0$ determines the critical
$\frac{f_{y}^{c}}{f_{x}^{c}}$, which clearly will remain much larger
than 0. This can be contrasted with the data for $I_{dc}=0.5$ and
0.8 mA for which $\frac{f_{y}}{f_{x}^{c}}$ decreases with decreasing
cutoff voltage, which could indicate that in this regime the
transverse pinning is not critical.

\begin{figure}
[ptb]
\begin{center}
\includegraphics[
natheight=9.635700in,
natwidth=6.199000in,
height=5.489in,
width=3.5405in
]%
{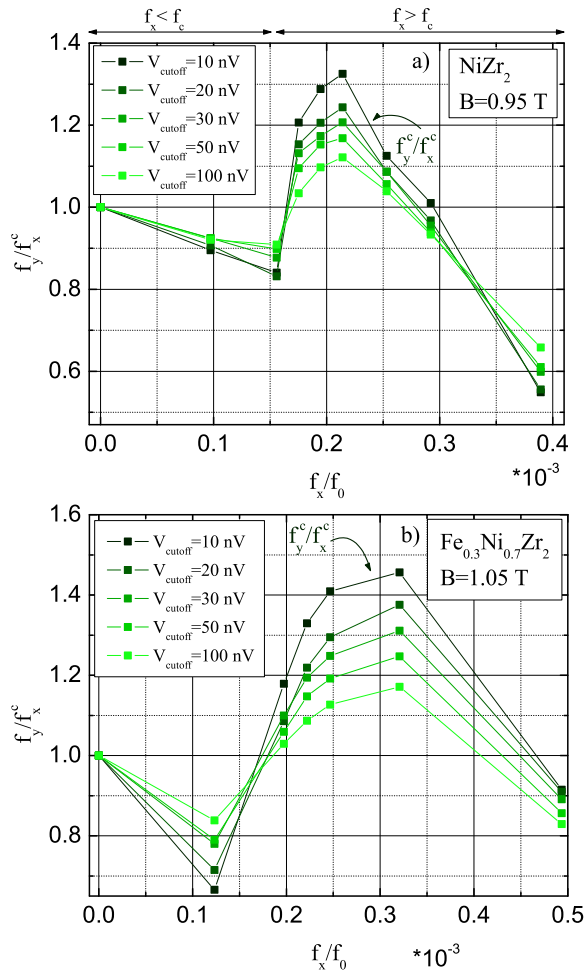}%
\caption{Ratio of the critical forces in the driven and the static case versus
the longitudinal dc force taken using different cutoff voltages for a) a
sample of NiZr$_{2}$ at B=0.95 T (main panel) and B=1.05 T (inset) b) a sample
of Fe$_{0.3}$Ni$_{0.7}$Zr$_{2}$ at B=1.05 T.}%
\label{fratiovsfdc}%
\end{center}
\end{figure}

In order to illustrate this critical behavior further, we choose
different cutoff voltages to extract the ratio
$\frac{f_{y}}{f_{x}^{c}}$ from the data shown in
Fig.(\ref{VvsFratio}) and plot this as a function of the dc
longitudinal force as shown in Fig.(\ref{fratiovsfdc}) for two
different samples.  As mentioned previously, the general behavior
observed with increasing dc force is an initial slight decrease of
$\frac{f_{y}}{f_{x}^{c}}$ followed by a strong increase reaching a
maximum at $f_{dc}=0.2\times10^{-3}f_{0}$ where $f_{0}$ is the
interaction force between two vortices separated by a distance
$\lambda$.  In the region below $f_{x}=0.16\times10^{-3}f_{0}$, the
longitudinal force is smaller than the longitudinal depinning force,
i.e. the ratio $\frac{f_{y}^{c}}{f_{x}^{c}}$ is dominated by the
transverse motion and the ratio should therefore be close to one.
However, the observed small initial decrease of
$\frac{f_{y}^{c}}{f_{x}^{c}}$ is attributable to the small
longitudinal dc component in the transverse direction due to our
small contact misalignment, which now helps depinning in the
transverse direction. In the peak region,
$0.16\times10^{-3}f_{0}<f_{x}<0.33\times10^{-3}f_{0}$, the
longitudinal force is now greater than the longitudinal depinning
force and an important enhancement of the transverse depinning force
is observed. This is also the region described above in which we
have a true critical transverse force. A strong dependence on the
choice of voltage cutoff is also observed, which signifies that the
effect of the longitudinal dc drive is not the same for all
transverse ac forces. We choose the curve with $V_{cutoff}=10~nV$ to
determine the critical transverse force
$\frac{f_{y}^{c}}{f_{x}^{c}}$ because, as discussed earlier, the
criticality of the transverse depinning transition is determine by
the behavior in the limit where $V_{cutoff}$ approaches zero.

The most striking overall feature is the very large magnitude of
$\frac{f_{y}^{c}}{f_{x}^{c}}$ measured here (between 0.55 and 1.33),
when compared to the numerical studies \cite{MoonPRL77, RyuPRL77,
OlsonPRB61, FangohrPRB63} (between 0.01 and 0.1). This much larger
magnitude of the normalized critical transverse depinning force in
our experimental weakly pinned system is consistent with the
numerical study in Ref. \cite{FangohrPRB63}, where they find the
critical forces ratio to be larger for weaker pinned simulated
samples. This increase in critical ratio for a decreasing
longitudinal pinning was attributed to a transverse depinning, which
is largely independent of the longitudinal depinning.
For a quantitative comparison, we use $f_{p}=A\left\vert \vec{J}%
_{c}\times\vec{B}\right\vert $, where $A$ is the area of the sample
perpendicular to the $B$ field, to obtain the pinning force per unit
length for our sample. This leads to $f_{p}=0.02f_{0}$, which means
it is 6 times less pinned than the weakest-pinned sample simulated
in Ref.\cite{FangohrPRB63}.  In addition, we obtain the critical
longitudinal depinning force for our system using the $V=10~nV$
cutoff to be $f_{c}=1\times10^{-4}f_{0}$, which is more than 200
times smaller than the longitudinal depinning force simulated in
Ref.\cite{FangohrPRB63}. These quantities confirm the weak-pinning
nature of our samples, which leads to the very large observed
critical transverse to longitudinal force ratio.

We have investigated experimentally the transverse dynamics of a vortex system
by measuring the resistance developed in a superconductor upon application of
an ac current in the transverse direction in the presence of a dc current in
the longitudinal direction.  We obtain values for the depinning force in the
driven case which are increased up to 33 \% with respect to the depinning
force in the static case, depending on the dc current used.  We attribute the
large magnitude of the critical force ratio in the driven and the static
cases found here to the weak-pinning properties of our samples.  We also
establish that the longitudinal drive suppresses vortex motion after
transverse depinning such that the vortices still feel some transverse pinning
at high velocity.

\begin{acknowledgments}
The authors acknowledge support from FQRNT and NSERC. \
Correspondence and requests for materials should be sent to
hilke@physics.mcgill.ca
\end{acknowledgments}

\bibliography{bibli}

\end{document}